\documentclass[aps,prc,floatfix,showpacs]{revtex4}
\usepackage{bm}
\usepackage{amsmath}
\usepackage{epsfig}
\usepackage{graphicx}
\usepackage{pdfpages}
\usepackage{color}   

\newcommand{\sat}{\mathrm{sat}}

\newcommand{\vac}{\mathrm{vac}}
\newcommand{\medium}{\mathrm{medium}}
\newcommand{\vacuum}{\mathrm{vacuum}}

\begin{document}

\title{Constraints on the in-medium nuclear interaction from chiral symmetry and Lattice-QCD}

\author{G. Chanfray } 
\affiliation{Univ Lyon, Univ Claude Bernard Lyon 1, CNRS/IN2P3, IP2I Lyon, UMR 5822, F-69622, Villeurbanne, France}
\author{H. Hansen} 
\affiliation{Univ Lyon, Univ Claude Bernard Lyon 1, CNRS/IN2P3, IP2I Lyon, UMR 5822, F-69622, Villeurbanne, France}
\author{J. Margueron} 
\affiliation{Univ Lyon, Univ Claude Bernard Lyon 1, CNRS/IN2P3, IP2I Lyon, UMR 5822, F-69622, Villeurbanne, France}

\date{\today}

\begin{abstract}
In this paper we discuss the combined effects on nuclear matter properties of the quark confinement mechanism in nucleon and of the chiral effective potential resulting from the spontaneous breaking of the chiral symmetry in nuclear matter. 
Based on the Nambu-Jona-Lasinio predictions, it is shown that the chiral potential acquires a specific scalar field cubic dependence, which contributes to the three-body interaction.
We also discuss the constraints induced by Lattice-QCD on the model parameters governing the saturation properties. We introduce the term "QCD-connected parameters" for these quantities. We demonstrate that chiral symmetry and Lattice-QCD provide coherent constraints on the in-medium nuclear interaction, suggesting a fundamental origin of the saturation mechanism.
\end{abstract}


\maketitle

\section{Introduction}\label{Intro}

Relativistic theories of nuclear matter initiated by Walecka and collaborators \cite{SerotWalecka1986,Walecka1997} attract a lot of interest for, at least, two reasons: i) this type of approach provides a very economical saturation mechanism and ii) a spectacular well-known success in predicting the correct magnitude of the spin-orbit potential since nucleons move in an attractive background scalar field and in a repulsive vector background field which contribute in an additive way (see a recent discussion for this specific point in Ref.~\cite{ChaMarg}). If the origin of the repulsive vector field can be safely identified as associated with the omega vector-meson exchange, the real nature of the attractive Lorentz scalar field has been  a controversial subject since there is no sharp scalar resonance with a mass of about 500-700~MeV, which would lead to a simple interaction based on a scalar particle exchange. More fundamentally the question of the very nature of these background fields has to be elucidated; in other words, it is highly desirable to clarify their relationship with the QCD condensates, in particular the chiral quark condensate $\left\langle \overline{q}q\right\rangle$, and more generally with the low energy realization of chiral symmetry which is spontaneously broken in the QCD vacuum and is expected to be progressively restored when the density increases. Indeed the microscopic origin of low-energy nuclear interaction properties is related to fundamental properties of the theory of the strong interaction (QCD) and should be implemented in the modeling of nuclear matter. 

To bridge the gap between relativistic theories of the Walecka type and approaches insisting on chiral symmetry, it has been proposed in Ref.~\cite{Chanfray2001} to identify the "nuclear physics" scalar sigma meson of the Walecka model at the origin of the nuclear binding, let us call it $\sigma_W$, with the chiral invariant $s=S-F_\pi$ field associated with the radial fluctuation of the chiral condensate $S$ around the "chiral radius" $F_\pi$, identified with the pion decay constant. In the present approach we take the point of view that the effective theory has to be formulated, as a starting point, in term
of the field $W$ associated with the fluctuations of the chiral quark condensate and parameterized as
\begin{eqnarray}
W&=&\sigma + i\vec{\tau}\cdot\vec{\pi}\equiv S\, U\equiv (s\, +\, F_{\pi})\,U\equiv (\sigma_W +\, F_{\pi})\,U\nonumber\\
&&\hbox{with}\qquad U(x)=e^{i\,{\vec{\tau}\cdot\vec{\phi}(x)}/{F_\pi}}.\label{REPRES}
\end{eqnarray}
The scalar field $\sigma$ ($S$) and pseudoscalar fields $\vec{\pi}$ ($\vec{\phi}$) written in cartesian (polar) coordinates  appear as the dynamical degrees of freedom and may deviate from the vacuum value, $\left\langle \sigma\right\rangle_{\vac}=\left\langle S\right\rangle_{\vac} = F_\pi\propto\left\langle \overline{q}q\right\rangle_{\vac}$. The sigma and the pion, associated with the amplitude  $s\equiv\sigma_W$ and phase fluctuations $\vec{\phi}$ of this condensate, are considered in our approach to be effective degrees of freedom. Their dynamics are governed by an effective chiral potential, $V\left(\sigma,\vec{\pi}\right)$, having a typical Mexican hat shape associated with a broken (chiral) symmetry of the QCD vacuum.

There is however  a well identified problem concerning the nuclear saturation with usual chiral effective theories \cite{Boguta83,KM74,BT01,C03}: independently of the particular chiral model, in the nuclear medium the value of $S$ ($\equiv S_{\medium}$) will be different from the one in vacuum ($\equiv S_{\vacuum}$, the minimum of  the vacuum effective potential represented by a "Mexican hat" potential). At $S_{\medium}$ the chiral potential has a smaller curvature : $V''(S_{\medium}) < V''(S_{\vacuum})$. This single effect results in the lowering of the sigma mass and destroys the stability, which is a problem for the applicability of such effective theories in the nuclear context. The effect can be associated with a $s^3$ tadpole diagram  generating attractive three-body forces destroying saturation even if the repulsive three-body force from the Walecka mechanism is present. The origin of this problem is most probably related to the fact that nucleons are not point particle, but in reality composite systems made of quarks. Hence the nucleon will react against the presence of the nuclear scalar field. This effect can be taken into account by introducing the nucleon response to the scalar field $s$, $\kappa_\mathrm{NS}=d^2M_N^*(s)/ds^2$ with the nucleon mass $M_N^*(s)$ defined in Eq.~\eqref{eq:mns}, which is the central ingredient of the quark-meson coupling model (QMC), introduced in the original pioneering work of P. Guichon \cite{Guichon1988} and successfully applied to finite nuclei with an explicit connection to the Skyrme force \cite{Guichon2004}. This effect associated with the polarization of the quark substructure in presence of the nuclear scalar field, will unavoidably generate  three-body forces  which may bring the needed repulsion. In practice this response or more precisely the nucleon scalar  susceptibility $\kappa_\mathrm{NS}$ generates a non-linear coupling of the scalar field to the nucleon or equivalently a decrease of  the scalar coupling constant with increasing density. Hence to achieve saturation, in a set of successive works devoted to the study of ordinary nuclear matter and neutron stars \cite{Chanfray2005,Chanfray2007,Massot2008,Massot2009,Massot2012}, we have complemented the relativistic chiral approach in such a way that the effect of the nucleon response is able to counterbalance the attractive chiral tadpole diagram to get good saturation properties, especially the correct curvature coefficient - the incompressibility modulus which is an empirical parameter defined at saturation density. All these aforementioned approaches were based on a chiral effective potential of the simplest linear sigma model 
with a Mexican hat shape of the following form
\begin{equation}
V_{\chi,\mathrm{L\sigma M}}(s)=\frac{1}{2}\,M^2_\sigma s^2\, +\,\frac{1}{2}\frac{M^2_\sigma -M^2_\pi}{ F_\pi}\, s^3\,+\,
\frac{1}{8}\,\frac{M^2_\sigma -M^2_\pi}{ F^2_\pi} \,s^4 \label{eq:VLSM}\, ,
\end{equation}
which displays a strong cubic tadpole term, also referred as the tadpole diagram~\cite{BT01,Ericson2007,Chanfray2008}. Indeed in order to get a correct description of the saturation properties it requires systematically a value of the dimensionless nucleonic response parameter, defined as (see also Eq.~\eqref{eq:coefC}), $$C\equiv \frac{\kappa_\mathrm{NS}\,F_\pi^2}{2 M_N},$$ to be larger than one \cite{Chanfray2005,Chanfray2007,Massot2008,Massot2009,Massot2012}. Such values are also required by the analysis of Lattice-QCD (LQCD) data on the chiral properties of the nucleon, with mass $M_N$, scalar charge $Q_S=\partial M_N/ \partial m$, and chiral susceptibility $\chi_N=\partial^2 M_N/ \partial m^2$  \cite{LTY03,LTY04,TGLY04,AALTY10} ($m$ is the current quark mass governing the explicit chiral symmetry breaking). Moreover in a recent work based on a Bayesian analysis with lattice data as an input \cite{Rahul}, we found that the response parameter is strongly constrained to a value $C\sim 1.4$ very close to the value where the scalar susceptibilities changes its sign: $C=1.5$.

The problem associated with this large value of $C$ is that it seems impossible to find a realistic confining models for the nucleon able to generate  $C$ larger than one. For instance in the MIT bag model used in the QMC scheme, one has $C_\mathrm{MIT}\simeq 0.5$. One possible reason for this discrepancy between models and phenomenological values of $C$ lies in the  use of the L$\sigma$M which  is probably too naive. Hence one should certainly use an enriched chiral effective potential from a model able to give a correct description of the low-energy realization of chiral symmetry in the hadronic world. A good easily tractable candidate is the Nambu-Jona-Lasinio (NJL) model. Indeed in Ref.~\cite{Chanfray2011}, referred as [NJLCONF] (NJL plus confinement) in the following, an explicit construction of the background scalar field was performed in the NJL model using a bosonization technique based on an improved derivative expansion valid at low (space-like) momenta~\cite{Chan}. Various confining interactions have been incorporated (quark-diquark string interaction, linear and quadratic confining interaction) on top of the NJL model which seem to be sufficient to generate saturation although the response parameters $C$ remain relatively small on the order of $C\sim  0.5$. The reason is that, for a given scalar mass, the NJL chiral effective potential generates a significantly smaller attractive tadpole diagram than the simplistic L$\sigma$M. We will discuss this point in more details  and demonstrate that the repulsive three-body force generating saturation, is not only determined by the nucleon response $C$ but also by the cubic term of the NJL potential, hereafter described by the new parameter $C_\chi$. The parameters $C$ and $C_\chi$ combine together in the three-body interaction. We will also demonstrate how a particular combination of $C$ and $C_\chi$ is constrained by lattice data \cite{LTY03,LTY04,TGLY04,AALTY10}, which constitutes one main result of this paper.

In this paper we mainly discuss the effect of the chiral effective potential, i.e., the contribution of the $C_\chi$ parameter, on the nuclear matter equation of state and on  the saturation mechanism, without explicitly specifying  the underlying nucleon confinement model. As mentioned above, very simple confining models have been already presented in [NJLCONF] and in a longer forthcoming paper referred as [NJLFCM] \cite{NJLFCM}, we will explicitly introduce an effective Hamiltonian inspired from the field correlator method (FCM) developed by Y. Simonov and collaborators \cite{Simonov1997,Tjon2000,Simonov2002a,Simonov2002,Simonov-light}. Modulo some ansatz prescription this approach allows us to generate simultaneously, at a semi-quantitative level,  a confining interaction with long distance ($r\gg T_g$)  behaviour $V(r)=\sigma_g \, r$, where the string tension $\sigma_g=0.18$~GeV$^2$,  together with an equivalent NJL model  with scalar interaction strength $G_1=120\pi\sigma_g T^4_g/(4N_cN_F)\sim 10$~GeV$^{-2}$ and cutoff $\Lambda\sim 1/T_g\sim 600$~MeV, where the gluon correlation length, $T_g=0.25$ to $0.3$~fm \cite{Digiacomo}, is itself related to the gluon condensate, $\mathcal{G}_2$, according to $T^2_g=9 \sigma_g/(\pi^3\mathcal{G}_2)$. Note that the string tension $\sigma_g$ and the gluon correlation length $T_g$ are two parameters measured in Lattice-QCD \cite{Digiacomo}.

\section{The NJL chiral confining model}

The general picture underlying our approach has been sketched in our previous papers (see, e.g., [NJLCONF]) and will be precised in our forthcoming work [NJLFCM]. It can be summarized as follows: nuclear matter is made of nucleons, themselves built from quarks and gluons which look like Y-shaped strings generated by a non perturbative confining force, with constituent quarks at the ends. These quarks acquire a large mass from the quark condensate, which is the order parameter associated with the spontaneous breaking of chiral symmetry in the QCD vacuum. When the density $n$ of nuclear matter increases, the QCD vacuum is modified by the presence of the nucleons: the value of the quark condensate decreases and the chiral symmetry is progressively restored. Hence what is usually called "the nuclear medium" can be seen as the original "vacuum shifted" by a lower value of the order parameter. The mass of the constituent quarks coincides with the in-medium expectation value, $M=\overline{\mathcal{S}}(n)$, of the chiral invariant scalar field $\mathcal{S}$, associated with the radial fluctuation mode of the chiral condensate. We define an "effective" or "nuclear physics" scalar field $s$ by rescaling the chiral invariant scalar field $\mathcal{S}$, according to: 
\begin{equation}
\mathcal{S} \equiv\frac{M_0}{F_\pi}\,S\equiv\frac{M_0}{F_\pi}\,\left(s+F_\pi\right) \quad\to\quad \frac{\partial}{\partial s}  = \frac{M_0}{F_\pi}\,\frac{\partial}{\partial \mathcal{S}}\label{Zfactor}
\end{equation}
where $M_0\sim 350$~MeV is the constituent quark mass in vacuum: $\overline{\mathcal{S}}(s=0)=M_0$. The vacuum expectation value of the "effective" scalar field, $\overline{S}=F_\pi$,  coincides  by construction with the value of the pion decay constant  ${F_\pi}$. The details of this construction are given in Ref.~\cite{Chanfray2011}. The important point is that its fluctuating piece, i.e., the  $s$ field, has to be identified with the usual "nuclear physics sigma meson" of relativistic Walecka theories, $\sigma_W$.

The nucleon is assumed to be described by an underlying model where constituent quarks (or diquarks) move in a confining interaction. In the previous [NJLCONF] work, ad-hoc confining potentials have been used on top of the NJL model generating the chirally broken vacuum. In the forthcoming longer paper [NJLFCM] the shape of this effective confining potential and the parameters of the equivalent NJL model will be obtained simultaneously in a way inspired from the field correlator method (FCM)\cite{Simonov1997,Tjon2000,Simonov2002a,Simonov2002,Simonov-light}. The nucleon mass will thus naturally depend on the scalar field whose expectation value, $M=\overline{\mathcal{S}}(n)$, is associated with the in-medium constituent quark mass, namely:
\begin{equation}
M^*_N(\mathcal{S}) = M_N + \, G_S\,  \left(\mathcal{S}-M_0\right)+3\, \frac{C_N}{M_0}\,
\left(\mathcal{S}-M_0\right)^2 +... .
\label{eq:expS}
\end{equation}
In passing we can notice that this approach is in spirit identical with the approach of Bentz and Thomas \cite{BT01} but with a different underlying picture of the nucleon; in this latter paper the nucleon was constructed from the same NJL model  as a bound quark-diquark state and the effect of confinement was taken into account through the presence of an infrared cutoff in the NJL loop integrals. We also used in our previous [NJLCONF] paper \cite{Chanfray2011} a simple quark-diquark NJL model but with confinement incorporated through a string interaction between the color antitriplet diquark state and the color triplet quark state as in a heavy $Q\overline Q$ meson.

The two dimensionless response parameters, $G_S$ which can be seen as  the scalar number of quarks in the nucleon, and the susceptibility  parameter $C_N$, only depend on the constituent quark mass and on the confining force, i.e., the confinement mechanism: 
\begin{equation}
G_S=\left(\frac{\partial M^*_N(\mathcal{S})}{\partial\mathcal{S} }\right)_{\mathcal{S}=M_0},\qquad
C_N=\frac{M_0}{6}\left(\frac{\partial^2 M^*_N(\mathcal{S})}{\partial\mathcal{S}^2 }\right)_{\mathcal{S}=M_0}.
\label{eq:gs}
\end{equation}
One important purpose of the present paper is to obtain phenomenological constraints on these two fundamental parameters that we will call "QCD-connected parameters", 
whereas our forthcoming paper \cite{NJLFCM} will provide a model calculation of these parameters in terms of $\sigma_g$ and $T_g$ within the FCM approach. 

\subsection{The NJL chiral effective potential}

In the following, we connect the expansion~\eqref{eq:expS} of the nucleon mass to previously published expansion in terms of the effective "nuclear physics" scalar field $s$~\cite{Chanfray2005,Chanfray2007,Massot2008,Massot2009, Chanfray2001,Massot2012,Rahul}, defined as:
\begin{equation}
s=\frac{F_\pi}{M_0}\left(\mathcal{S}-M_0\right) \, .
\end{equation}
We have the following expansion of the nucleon mass:
\begin{eqnarray}
M^*_N(s)&=& M_N + g_S s + \frac{1}{2}\kappa_\mathrm{NS} s^2 + \mathcal{O}(s^3) \,=\, M_N\left(1 + \frac{g_S F_\pi}{M_N}\frac{s}{F_\pi} + C \left(\frac{s}{F_\pi}\right)^2 +\dots \right)\, , \label{eq:mns}\\
\hbox{with:}&& g_S=\frac{M_0}{F_\pi} G_S\,,\qquad C\equiv \frac{\kappa_\mathrm{NS} F_\pi^2}{2 M_N}=\frac{3 M_0}{M_N}C_N\,.\label{eq:coefC}
\end{eqnarray}
Consequently the in-medium nucleon mass mainly depends on two effective dimensionless QCD-connected parameters, the scalar nucleon coupling constant, $g_S$, and the dimensionless scalar nucleon susceptibility, $C\equiv\kappa_\mathrm{NS}\,F_\pi^2/2 M_N$, which embeds the influence of the internal nucleon structure or said differently the response of the nucleon to the nuclear scalar field. Notice that the response parameter $C$ used in our previous work is numerically close to the QCD-connected susceptibility  parameter $C_N$. Its presence generates a decreasing density dependence of the in-medium scalar coupling constant, $g^*_S(s)=\partial M^*_N/\partial s=g_S + \kappa_\mathrm{NS}\, s +.. $, corresponding to a progressive decoupling of the nucleon from the chiral condensate, which is  an essential ingredient of the  saturation mechanism (recall that $s$ is a negative quantity varying between zero in the vacuum to $- F_\pi$ at full chiral restoration).

The nuclear matter energy density as a functional of the scalar field ${\mathcal{S}}$ or the $s$ field  is given by 
\begin{equation}
\varepsilon_0=\int\,\frac{4\,d^3 k} {(2\pi)^3} \,\Theta(p_F - k)\,\left(\sqrt{k^2+M^{*2}_{N}({s})}
\,-\,M_{N,\mathrm{vac}}\right)
\,+\,V_\chi({s}) \,+\,\varepsilon_{\omega+\rho}\,+\,\varepsilon_\mathrm{Fock}\,+\,\varepsilon_\mathrm{pion-nucleon\, loops} ,
\end{equation}
where only the scalar field contribution at the Hartree level together with the kinetic energy are explicitly written, while omega and rho meson exchanges, Fock terms and pion-nucleon loops (or correlation energy in the terminology of Ref.~\cite{Chanfray2007}) can be incorporated as well according to Refs.~\cite{Chanfray2007,Massot2008,Massot2009}. Note that $V_\chi({s})$ is the chiral effective potential which is expressed in the L$\sigma$M by Eq.~\eqref{eq:VLSM}. 

Let us now consider the case of  the NJL model defined by the Lagrangian:
\begin{eqnarray}
{\cal L}&=& \overline{\psi}\left(i\,\gamma^{\mu}\partial_\mu\,-\,m\right)\,\psi\,+\,\frac{G_1}{2}\,\left[\left(\overline{\psi}\psi\right)^2\,+\
\left(\overline{\psi}\,i\gamma_5\vec\tau\,\psi\right)^2\right]\nonumber\\
& &-\,\frac{G_2}{2}\,\left[\left(\overline{\psi}\,\gamma^\mu\vec\tau\,\psi\right)^2\,+\,
\left(\overline{\psi}\,\gamma^\mu\gamma_5\vec\tau\,\psi\right)^2\,+\,\left(\overline{\psi}\,\gamma^\mu\,\psi\right)^2\right]. \label{LNJL}
\end{eqnarray}

It depends on four parameters: the coupling constants $G_1$ (scalar), $G_2$ (vector), the current quark mass $m$ and a (noncovariant) cutoff parameter $\Lambda$. Three of these parameters ($G_1$, $m$, and $\Lambda$) are adjusted to reproduce the pion mass, the pion decay constant and the quark condensate. For $G_2$ we consider different scenarios: $G_1=G_2$ and $G_2=0$. We refer the reader to [NJLCONF] and [NJLFCM] for more details. Using path integral techniques and after a chiral rotation of the quark field, it can be equivalently  written in a semi-bozonized form involving a pion field $\vec{\phi}$ embedded in the unitary operator $U=\xi^2=exp(i\,\vec{\tau}\cdot\vec{\phi}(x)/{F_\pi})$, a scalar field, ${\cal S}$, a vector field, ${V}^\mu$, and an axial-vector field,  ${A}^\mu$. It has the explicit form given in Eqs. (2, 7-11) of Ref.~\cite{Chanfray2011}. Subtracting the vacuum expectation values, the chiral effective potential can be expressed as:
\begin{equation}
V_{\chi,\mathrm{NJL}}(s)=-2N_c N_f\,\big(I_0(\mathcal{S})\,-\,I_0(M_0)\big) \,+\,\frac{\left(\mathcal{S}-m\right)^2 -\left(M_0 - m\right)^2}{2\,G_1}.\label{VNJL}
\end{equation}
The quantity, $-2N_c N_f\,I_0(\mathcal{S})$, is nothing but the total (in-medium) energy of the Dirac sea of constituent quarks with the NJL loop integral $I_0(\mathcal{S})$ given hereafter. The vacuum constituent quark mass $M_0$  corresponds to the minimum of the chiral effective potential, i.e., $V'_{\chi,\mathrm{NJL}}(s=0)=0$, where $V'$ is the derivative with respect to the scalar field $s$. It is consequently the solution of the gap equation
\begin{equation}
M_0 = m\,+\,4N_c N_f M_0\,G_1\,I_1(M_0),    \label{GAP}
\end{equation}
where $I_1(M_0)$ is another NJL loop integral given in the set of equations below 	
\begin{eqnarray}
&& I_0(\mathcal{S})=\int_0^\Lambda \frac{d{\bf p}}{(2\pi)^3}\,E_p(\mathcal{S}),\quad	I_1(\mathcal{S})=\int_0^\Lambda \frac{d{\bf p}}{(2\pi)^3}\,\frac{1}{2\,E_p(\mathcal{S})},\nonumber\\
&& I_2(\mathcal{S})=\int_0^\Lambda \frac{d{\bf p}}{(2\pi)^3}\,\frac{1}{4\,E^3_p(\mathcal{S})},	\quad J_3(\mathcal{S})=\int_0^\Lambda\frac{d{\bf p}}{(2\pi)^3}\,\frac{3}{8\,E^5_p(\mathcal{S})} \, , \label{PARAM1}
\end{eqnarray}
where $E_p(\mathcal{S})=\sqrt{\mathcal{S}^2 + p^2}$.

\subsection{Effective chiral potential expanded in the $s$ field}

For a comparison with usual RMF model using the L$\sigma$M chiral effective potentials of Eq.~\eqref{eq:VLSM} or equivalently non-linear sigma couplings, we expand the effective potential to third order in $s$ as:
\begin{equation}
V_{\chi,\mathrm{NJL}}({s})=V_{\chi}(0)+V'_{\chi}(0)\,{s}+\frac{1}{2}V''_{\chi}(0)\, {s}^2 +\frac{1}{6}V'''_{\chi}(0)\, {s}^3 +... .
\end{equation}	
An explicit calculation of the derivatives of the potential yields
\begin{equation}
V_{\chi,\mathrm{NJL}}(s)= \frac{1}{2}\,M^2_\sigma\, {s}^2\, +\,\frac{1}{2}\,\frac{M^2_\sigma -M^2_\pi}{ F_\pi}\, {s}^3\,\big(1\,-\,C_{\chi,\mathrm{NJL}}\big) +...,
\label{eq:vchiNJL}
\end{equation}
where $F_\pi$ is the pion decay constant and $M_\pi=\sqrt{m M_0/G_1 F^2_\pi}$, the canonical pion mass calculated in the bosonized NJL model. The effective sigma mass $M_\sigma$ (considering the axial-pion mixing) is defined as
\begin{equation}
M^2_\sigma=4\,M^2_0 \, \frac{f^2_\pi}{F^2_\pi} + \,M^2_\pi, \qquad
\text{with: }\quad f^2_\pi = \frac{F^2_\pi}{1-4G_2 F^2_\pi} 
\label{eq:msigma}
\end{equation}
(where the second relation is obtained in the NJL model \cite{Chanfray2011})
and $C_{\chi,\mathrm{NJL}}$ is a specific NJL parameter: 
\begin{equation}
C_{\chi,\mathrm{NJL}}=\frac{2}{3}\,\frac{M^2_0\,J_3(M_0)}{I_2(M_0)}.    
\end{equation}
This form of the NJL chiral effective potential deviates from the original L$\sigma$M, see Eq.~\eqref{eq:VLSM}, through the presence of the model dependent parameter $C_{\chi,\mathrm{NJL}}$ whose net effect is to decrease the attractive cubic tadpole term of the L$\sigma$M. The use of this $C_{\chi,\mathrm{NJL}}$ parameter is particularly convenient, since taking $C_{\chi,\mathrm{NJL}}=1$ is equivalent to the absence of the tadpole diagram as in the case of the QMC model \cite{Guichon1988,Guichon2004}.  

\begin{figure}
\centering
\includegraphics[width=0.8\textwidth,angle=0]{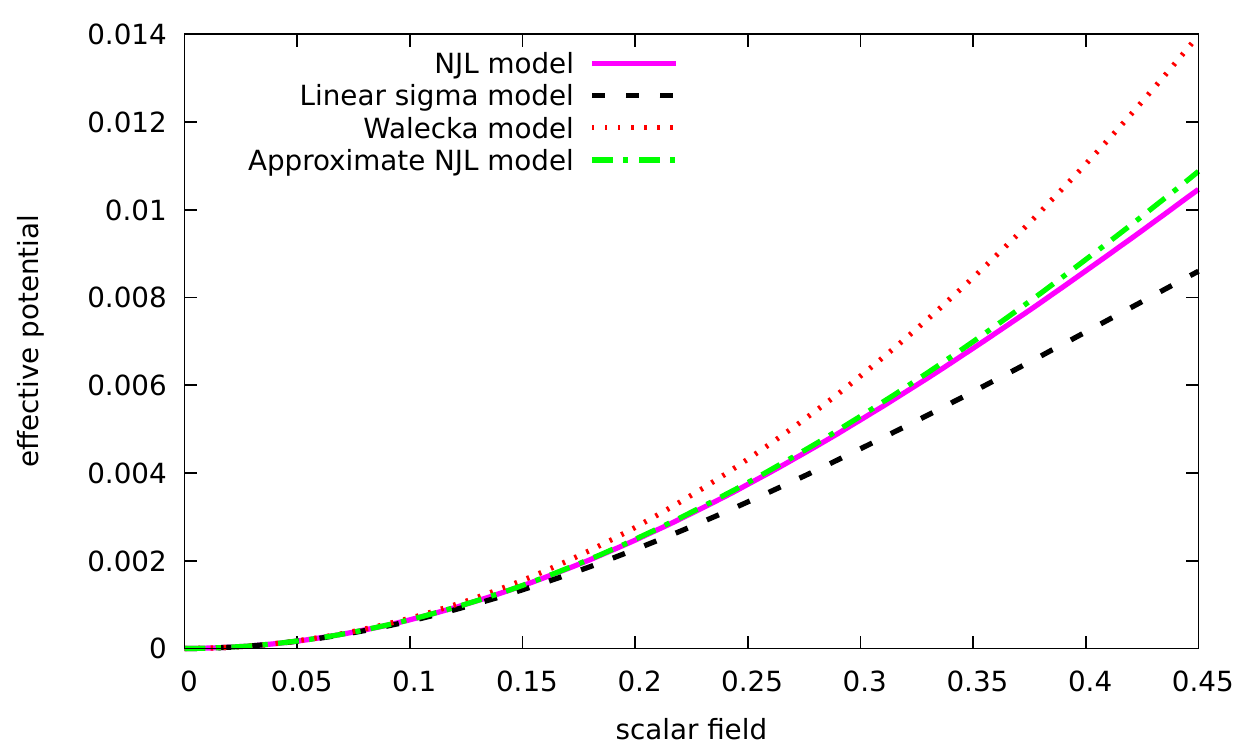}
\caption{Effective potential (in units of the string tension $\sigma^2$, with $\sigma_g=0.18$~GeV$^2$) plotted against $\vert s\vert/F_\pi$ for the NJL model (full line), L$\sigma$M (dashed line) and original Walecka model that is limited to the quadratic term (dotted line), for a given effective sigma mass $M_\sigma=716.4$~MeV. Also shown is the approximate form of the NJL potential when limited to the cubic term in the scalar field $s$ expansion~\eqref{eq:vchiNJL} (dot-dashed line). The effect of the $s^3$ term is well seen when comparing to the Walecka model. 
Note that the approximate expansion~\eqref{eq:vchiNJL}  is almost identical to the exact  NJL potential.}
\label{fig:f2}
\end{figure}

In the absence of vector interaction ($G_2=0$), for typical value of FCM parameters, $\sigma_g=0.18$~GeV$^2$, $T_g=0.286$~fm, one obtains  $G_1 = 12.514$ GeV$^{-2}$. The NJL cutoff behaves necessarily as $\Lambda\sim 1/T_g$  but there is a certain arbitrariness in setting its precise value: we take  $\Lambda= 0.604$~GeV. Taking $m = 5.8$~MeV this enables us to obtain reasonable values for the pion decay constant,  $F_\pi=91.9$~MeV, the pion mass $M_\pi=140$~MeV, and the quark condensate  $\langle \bar{q} q\rangle=-(241.1\,$MeV$)^3$.  The resulting vacuum constituent quark mass, effective sigma mass and $C_\chi$ parameter  are $M_0= 356.7$~MeV,  $M_\sigma=716.4$~MeV and  $C_{\chi,\mathrm{NJL}}= 0.488$. Fig.~\ref{fig:f2} shows that the approximate expansion~\eqref{eq:vchiNJL} reproduces very well the exact NJL potential. Comparing L$\sigma$M with NJL scalar potential in Fig.~\ref{fig:f2}, one sees that the attractive tadpole term is larger in the case of L$\sigma$M. The effect of the parameter $C_{\chi,\mathrm{NJL}}$ is then to reduce the attractive tadpole diagram and make the scalar potential more repulsive. Using another parameter set, $G_1 = 7.705$ GeV$^{-2}$, $\Lambda= 0.740$~GeV and $m = 3.5$ MeV, compatible with the $\pi-a_1$ mixing with $G_2=G_1$ as suggested by the FCM~\cite{Simonov1997,Tjon2000,Simonov2002a,Simonov2002,Simonov-light},  one obtains $M_0= 365.3$~MeV and a smaller value of  $C_{\chi,\mathrm{NJL}}= 0.43$  but the reduction of the tadpole diagram is still significant. 

In the following, we set $C_\chi\equiv C_{\chi,\mathrm{NJL}}$ and $V_\chi\equiv V_{\chi,\mathrm{NJL}}$ for simplicity.

\subsection{Impact on nuclear matter properties}

At the Hartree approximation (RMF), the scalar field minimizing the total energy is the solution of the following self-consistent equation of motion:
\begin{equation}
V'_\chi({s})= - g^*_S({s}) n_s\qquad \hbox{with}\qquad n_s =4\int_0^{k_F}\!\!\frac{d{\bf k}}{(2\pi)^3}\frac{M^*_N(s)}{\sqrt{ M^{*2}_N({s})+k^2}},
\label{GAPM}
\end{equation}
where $V'_\chi({s})$ is the derivative of the Mexican hat chiral effective potential, with respect to the scalar field $s$. This equation constitutes an in-medium modified gap equation whose solution is controlled by the nucleonic scalar density $n_s$.

To second order in ${s}/F_\pi$  or equivalently to second order in the scalar density $n_s$, the in-medium gap equation can be formally solved with the result: 
\begin{eqnarray}
\overline{s}&=&-\frac{g_S}{M^2_\sigma}\,n_s\,+\,\frac{g_S}{M^4_\sigma}\,\left(\kappa_\mathrm{NS}\,-\,\frac{g_S\,V'''_{\chi}(0)}{2\,M^2_\sigma}\right)\,n^2_s\nonumber\\
&=&-\frac{g_S}{M^2_\sigma}\,n_s\,+\,\frac{g_S}{M^4_\sigma}\,
\left(\frac{2\,M_N}{F^2_\pi}\, C\,-\,\frac{3\, g_S}{2\,F_\pi}\,\frac{M^2_\sigma\,-\,M^2_\pi}{M^2_\sigma}\,(1\,-\,C_\chi)\right)\,n^2_s\nonumber\\
&=&-\frac{g_S}{M^2_\sigma}\,n_s\,+\,\frac{g^2_S}{M^4_\sigma\,F_\pi}\,\left(2\, \Tilde{C}_s\,-\,\frac{3}{2}\right)\,n^2_s\qquad
\hbox{with}\qquad \Tilde{C}_s \simeq\frac{M_N}{g_S\,F_\pi}\,C +\,\frac{3}{4}\,C_\chi .
\end{eqnarray}
For a qualitative discussion, we have supposed $M_\pi\ll M_\sigma$ to get the approximate expression $\Tilde{C}_s$. 

The scalar field contribution to the energy per nucleon is defined as $E_s/A=V_\chi(s)/n+M_N^*(s)-M_N$. To leading order in density, its contribution is defined as $E^{(2b)}$, which reads
\begin{equation}
\frac{E^{(2b)}}{A}=-\frac{g^2_S}{M^2_\sigma}\,n_s\,+\,\frac{1}{2}\,\frac{g^2_S}{M^2_\sigma}\,\frac{n^2_s}{n}
=-\frac{1}{2}\,\frac{g^2_S}{M^2_\sigma}\,n\,+\,\frac{1}{2}\,\frac{g^2_S}{M^2_\sigma}\,\frac{\left(n_s - n\right)^2}{n}.
\label{eq:e2b}
\end{equation}
In the first expression of Eq.~\eqref{eq:e2b}, we have separated the effect of the scalar self-energy of the nucleon and the contribution of the effective potential at leading order in the densities $n$ and $n_s$. In the second form, we display explicitly  the  term proportional to $(n_s-n)^2$,  corresponding to  an effective repulsive three-body force, which is exactly the Walecka saturation mechanism when the omega is added. This contribution, which survives for an point-like nucleon,  is proportional to the square of the nucleon momentum. This is the so-called Z graph associated with the excitation of $N\overline{N}$ pairs \cite{Birse95,Wallace}.

To second order in density $E_s/A$ provides an effective three-body contribution to the energy per nucleon:
\begin{equation}
\frac{E^{(3b)}}{A}\simeq\frac{g^2_S}{2\,M^4_\sigma}\,\left(\kappa_\mathrm{NS}\,-\,\frac{g_S\,V'''_{\chi}(0)}{3\,M^2_\sigma}\right)\,n^2_s\,
=\,\frac{g^3_S}{\,M^4_\sigma\,F_\pi}\,\left(2\, \Tilde{C}_3\,-\,1\right)\,n^2_s\qquad
\hbox{with}\qquad \Tilde{C}_3 \simeq\frac{M_N}{g_S\,F_\pi}\,C +\,\frac{1}{2}\,C_\chi.
\label{eq:THREEBOD}
\end{equation}
We can recover Eq.~(44) of Ref.~\cite{Ericson2007} with $C_\chi=0$. 

We now give a qualitative discussion of the influence of the three parameters $g_S$, $\kappa_\mathrm{NS}$ and $V'''_{\chi}(0)$ or equivalently $g_S$, $C$ and $C_\chi$, taking various works as illustrative examples.\\

If we ignore both the response of the nucleon, i.e., $\kappa_\mathrm{NS}=0$ (or $C=0$), and the contribution of the tadpole diagram to the chiral potential, i.e., $V'''_{\chi}(0)=0$ (or $C_\chi=1$), we recover the original Walecka model since the three-body contribution~\eqref{eq:THREEBOD} is absent and the saturation mechanism is associated with the Z graph alone, see Eq.~\eqref{eq:e2b}.  It is known that in this case saturation requires a large $g_S/M_\sigma$ value, which implies a large repulsion induced by $g_\omega/m_\omega$ in order to obtain the empirical value of the binding energy. 
As a consequence one gets a much too large incompressibility modulus $K_{\sat}$. One possibility to cure this problem is to introduce density dependent coupling constants~\cite{Typel,Vandalen}.

In the QMC model originally proposed in Ref.~\cite{Guichon1988} and providing a successful phenomenology \cite{Guichon2004}, the response of the nucleon is incorporated, but without explicit connection with the chiral status of the scalar field. Hence no tadpole diagram is considered, i.e., $V'''_{\chi}(0)=0$ or $C_\chi=1$. The original QMC model is formulated in the MIT bag model, yielding $C\sim 0.5$ and $E^{(3b)} \propto 2 \Tilde{C}_3-1=2 C\sim 1$ which turns out to be sufficient to bring the needed repulsion to get nuclear saturation with a correct incompressibility modulus, although this approach does not satisfy chiral symmetry requirements.

Soon after the first version of the relativistic Walecka model, it has been realized \cite{Boguta83,KM74,BT01,C03} that in relativistic theory with a mexican hat-like effective potential, the contribution of the Walecka $Z$ graph  is not large enough to stabilize nuclear matter against the effect of the attractive tadpole diagram. This is the typical situation of the original L$\sigma$M where $V'''_{\chi}(0)$ is large and positive ($C_\chi\sim 0$) and even of the NJL model ($C_\chi < 1$) where the response of the nucleon is ignored, i.e., $C=0$. Some phenomenological approaches, such as the so-called NL3 model~\cite{NL3}, have introduced self-interactions of the scalar field in the form of an effective potential but without connection to chiral symmetry. In particular a repulsive cubic term, i.e., $V'''_{\chi}(0)<0$, is introduced in this model. From Table II of Ref.~\cite{NL3}, one can obtain the equivalent $C_\chi \sim 1.47$ parameter, which corresponds to  $\Tilde{C}_3 \sim 0.74$. One can thus re-interpret the original NL3 model with a negative value of the $c_2$ parameter (see table II of Ref.~\cite{NL3}) as a way to simulate in an effective way the nucleon response with $C\sim 0.74$. The way the non-linear potential has been introduced in the NL3 model was pragmatic, but it can now be understood in a more fundamental approach.

\section{Constraining the chiral confining potential by Lattice-QCD}

In this section, we connect the in-medium properties of the nucleon mass defined by Eq.~\eqref{eq:mns} with the Lattice-QCD calculations performed in vacuum ($s=0$). For this reason, the nucleon mass will be noted in the following $M_N(s)$ (without the $^*$). The derivatives of the nucleon mass could however be obtained, on the one hand, from the derivatives of the nucleon mass~\eqref{eq:mns} taken at $s=0$ and providing $g_S$ and $\kappa_\mathrm{NS}$, and, on the other hand, from the Lattice-QCD calculations.

Within an underlying microscopic confining model for the nucleon, i.e., [NJLCONF] and [NJLFCM], generating the quark core wave functions, the axial charge, the  $\pi NN$ coupling constant and the $\pi NN$ form factor can be obtained, allowing the calculation of  the  pion cloud contribution (pion self-energy) to the in-medium nucleon (and Delta resonance) mass, as in the Cloudy Bag model \cite{CBM} or similar approaches using an alternative confinement potential \cite{Jena97}. The pion contribution to the nucleon mass is expressed as
\begin{equation}
\Sigma^{(\pi)}({M;m})=-{3\over 2}\left({g_A(M)\over 2 F_\pi(M)}\right)^2\int{d{\bf q}\over 
(2\pi)^3} {\bf q}^2 v^2({\bf q}; M)\left(\frac{1}{\omega_q}\frac{1}{\omega_q+\epsilon_{N {\bf q}}}
+\frac{32}{25}\frac{1}{\omega_q}\frac{1}{\omega_q+\epsilon_{\Delta{\bf
q}}}\right) \, , \label{eq:SELFENERGY}
\end{equation}
with $\omega_q=\sqrt{q^2 +M^2_\pi(M)}$ and $M^2_\pi(M)={m M}/{G_1 F^2_\pi(M)}$, the other quantities being defined in Eq. (22) of Ref.~\cite{Chanfray2007}. Here the various quantities such as $M^2_\pi(M)$ are in-medium quantities  where the vacuum constituent quark mass $M_0$ is replaced by $M=\overline{\mathcal{S}}$ (see Eq. (34) of Ref.~\cite{Chanfray2011} and the text just before). Thus in this framework, the nucleon mass is split according to:
\begin{equation}
M_N({s})\equiv M_N(M; m)=M_N^\mathrm{core}(M)+\Sigma^{(\pi)}(M ;m)\label{eq:SPLIT}\, .
\end{equation}

\subsection{Nucleon response and its chiral properties}

The derivatives of the nucleon mass with respect to the constituent quark mass gives the response parameters, which are defined in Eq.~\eqref{eq:gs}, i.e., $G_S=\partial M_N(M;m)/\partial M$ and $C_N=(1/6)\partial^2 M_N(M;m)/\partial M^2$, where the derivatives are taken at $M=M_0(m)$, i.e. $s=0$. To benefit from the lattice data, we can relate them to two chiral properties of the nucleon, the scalar charge, $Q_S=\partial M_N(M=M_0(m);m)/ \partial m$, and the chiral susceptibility, $\chi_N=\partial^2 M_N(M=M_0(m);m)/ \partial m^2$. All what we need for this calculation are the derivatives of the constituent quark mass with respect to the current quark mass. 
These derivatives are obtained from the NJL model and read ([NJLFCM]):
\begin{equation}
\left(\frac{\partial M_0}{\partial m}\right) =
\left(\frac{M^2_\pi}{m}\right)\,\frac{M_0}{M^2_\sigma},\qquad
\left(\frac{\partial^2 M_0}{\partial m^2}\right)\approx 
-\,\left(\frac{M^2_\pi}{m}\right)^2\, \frac{3 M_0}{M^4_\sigma}\, \,\,(1-C_\chi).
\label{eq:derM0}
\end{equation}
where in the second expression a correction factor of order $M^2_\pi/M^2_\sigma$ has been neglected. We now note that in Eq.~\eqref{eq:SPLIT} the current quark mass appears explicitly only in the pionic self-energy $\Sigma^{(\pi)}$. It appears also implicitly through the dependence of the constituent quark mass upon the current quark mass. Hence the scalar charge, $Q_S$ \cite{Ericson2007}, receives two different contributions: 
\begin{eqnarray}
Q_S &=&\frac{\partial M_N}{\partial m}=\frac{\partial M_N}{\partial M}\left(\frac{\partial M_0}{\partial m}\right) + \frac{d\Sigma^{(\pi)}({M_0 ;m})}{d m}=\frac{F_\pi}{M_0}g_S\left(\frac{M^2_\pi}{m}\right)\frac{M_0}{M^2_\sigma}
+\left(\frac{M^2_\pi}{m}\right) \frac{d\Sigma^{(\pi)}(M_0 ;m)}{d M^2_\pi}\nonumber\\
&= &\left(\frac{M^2_\pi}{m}\right)\, \frac{F_\pi\,g_S}{M^2_\sigma}+\left(\frac{M^2_\pi}{m}\right) \frac{d\Sigma^{(\pi)}}{d M^2_\pi}
\equiv Q^{(s)}_S+Q^{(\pi)}_S,
\label{eq:QS}
\end{eqnarray}
where we have employed the relation~\eqref{Zfactor}. The second term, $Q^{(\pi)}_S$, is referred as the pion cloud contribution. It is obtained by taking only the linear quark mass dependence appearing in $M^2_\pi= M_0 m/G_1 F^2_\pi$, thus ignoring all the implicit $m$ dependencies through the $M$ dependence of $M_\pi$, $g_A$, $F_\pi$ and the form factor. We refer the first term, $Q^{(s)}_S$, as the scalar field contribution despite it contains the implicit $m$ dependence of the pionic self-energy. In effect $Q^{(s)}_S$ receives itself two separate contributions: 
\begin{equation}
Q^{(s)}_S=  \frac{\partial M^\mathrm{core}_N}{\partial M}\left(\frac{\partial M_0}{\partial m}\right)\,+\,  \frac{\partial\Sigma^{(\pi)}}{\partial M}\left(\frac{\partial M_0}{\partial m}\right)\equiv \left(\frac{M^2_\pi}{m}\right)\, \frac{F_\pi\,g_S}{M^2_\sigma}.\label{eq:SCALQ}
\end{equation}
The second contribution contains the implicit $m$ dependence of the pion self-energy coming from the $M$ dependence of  the various quantities ($F_\pi(M)$, $M_\pi(M)$, $g_A$, form factor)  through the $m$ dependence of the constituent quark mass  taken at its vacuum  value $M_0$. Regarding this specific point it is generally assumed that the pion properties  are  protected by  chiral symmetry and this is what we find in the model developped in [NJLFCM] where the pion mass  displays a remarkable stability for a large domain of the constituent quark mass or equivalently of the nuclear  scalar field $s$. As a consequence the induced effect on $g_S$ is extremely small.  However the combined effect of the modification of the nucleon size and of the pion decay constant might induce a more important  correction  on the $\pi NN$ vertex $(g_A v(q)^2/2 F_\pi)^2$ but we do not consider this effect which certainly requires a more detailed study. It follows for Eq.~\eqref{eq:QS} that:
\begin{equation}
m\,\frac{\partial M_N}{\partial m}=F_\pi\, g_S\,\frac{M^2_\pi}{M^2_\sigma}\,+\, M^2_\pi\,
\frac{d\Sigma^{(\pi)}({M_0 ;m})}{d M^2_\pi}\equiv\sigma_N^{(s)}\,+\,\sigma_N^{(\pi)}\equiv \sigma_N.
\end{equation}
Hence we recover the nucleon sigma term. This result is just the expression of the  Feynman-Hellman theorem. This light quark sigma term has been abundantly discussed in our previous papers \cite{Chanfray2007,Massot2008,Chanfray2011}. Using a dipole $\pi NN$ form factor with cutoff $\Lambda=1$~GeV, the pionic contribution to the sigma term was found to be $\sigma_N^{(\pi)}=21.5$~MeV \cite{Chanfray2007} and a pionic self-energy $\Sigma^{(\pi)}(M;m)=420$~MeV. The value of the non pionic contribution was found to be $\sigma_N^{(s)}\sim 29$~MeV \cite{Chanfray2007} to get a total sigma term $\sigma_N=50.5$~MeV. Evidently the relative weight of the two contributions may be altered by the precise values of the parameters, but according to our model FCM calculation \cite{NJLFCM} and from  the lattice data constraints discussed below, this modification of the relative weight should be rather moderate and the value of the sigma term and its repartition is a rather strong constraint on the nucleon modelling.

For the scalar susceptibility one obtains from Eq.~\eqref{eq:QS}, ignoring again higher order correction $M^n_\pi/M^n_\sigma$ 
\begin{eqnarray}
\chi_N &=&\frac{\partial^2 M_N}{\partial m^2}=\frac{\partial M_N}{\partial M}\left(\frac{\partial^2 M_0}{\partial m^2}\right)+\frac{\partial^2 M_N}{\partial M^2} \left(\frac{\partial M_0}{\partial m}\right)^2 + \frac{\partial}{\partial M}\left(\frac{d\Sigma^{(\pi)}}{dm}\right)\left(\frac{\partial M_0}{\partial m}\right)+\frac{d^2\Sigma^{(\pi)}}{dm^2}\nonumber\\
&=& -\,\left(\frac{M^2_\pi}{m}\right)^2\,\frac{3\,g_S\,F_\pi}{M^4_\sigma}\,(1-C_\chi)\,+\,
\left(\frac{M^2_\pi}{m}\right)^2 \kappa_\mathrm{NS}\, \frac{F^2_\pi}{M^4_\sigma} +\left(\frac{M^2_\pi }{m}\right)^2\frac{1}{M^2_\sigma}\frac{d}{d M^2_\pi}
\left(M_0\,\frac{\partial \Sigma^{(\pi)}}{\partial M}\right) +\frac{d^2\Sigma^{(\pi)}}{dm^2}\nonumber \\
&\equiv & \chi^{(s)}_N+\chi^{(s \pi)}_N\,+\,\left(\frac{M^2_\pi}{m}\right)^2 \frac{d^2\Sigma^{(\pi)}(M_0 ;m)}{d (M^2_\pi)^2}
\equiv  \chi^{(s)}_N+\chi^{(s \pi)}_N\,+\,\chi^{(\pi)}_N\, ,\label{eq:SUSC}
\end{eqnarray}
where we have used Eq.~\eqref{eq:derM0}. One can split the scalar susceptibility into a non pionic ($\chi^{(s)}_N$), a mixed scalar field-pionic ($\chi^{(s\pi)}_N$) and a purely pionic ($\chi^{ (\pi)}_N$) piece. The first two contributions in the second line of Eq.~\eqref{eq:SUSC} with $M^2_\sigma\sim4 M^2_0$ (considering small $M_\pi$ and $G_2$ in Eq.~\eqref{eq:msigma}), gives $\chi^{(s)}_N$ as:
\begin{equation}
\chi^{(s)}_N=-\,\left(\frac{M^2_\pi}{m}\right)^2\,\frac{F_\pi\,g_S}{M^4_\sigma}\,\left(3\,-\,2\,\tilde{C}_L \right)\quad\hbox{with}\quad
\tilde{C}_L = \frac{M_N}{g_S\,F_\pi}\,C\,+\,\frac{3}{2}\,C_\chi.
\label{eq:SCALS}
\end{equation}
As for the case of $g_S$, the nucleon susceptibility $\kappa_\mathrm{NS}$ may receive a contribution from the pion-self-energy; again the contribution to the dimensionless $C$ parameter is very small if the vertex correction is omitted. The mixed scalar field-pionic susceptibility originating from the scalar field (i.e, the constituent quark mass) dependence of the pionic self-energy,
\begin{equation}
\chi^{(s \pi)}_N=-\left(\frac{M^2_\pi}{m}\right)^2
\,\frac{F_\pi}{M^2_\sigma}\,
\frac{\partial}{\partial s}
\left( \frac{\sigma_N^{(\pi)}}{M^2_\pi}\right), 
\label{eq:CHINS}
\end{equation}
was ignored in our previous works. Using a sharp cutoff in the expression of the nucleon pionic self-energy, it can be shown analytically that this term is negligible compared to the other contributions to the susceptibility.

In view of the comparison with lattice QCD result it is very important to notice that the chiral susceptibility is governed by the particular combination:
\begin{equation}
\chi_N \sim \left(3\,-\,2\,\tilde{C}_L \right) \, .
\label{eq:xin}
\end{equation}
to be compared with the particular combination entering the expression of the three-body repulsive contribution~\eqref{eq:THREEBOD} to the binding energy per nucleon: 
\begin{equation}
\frac{E^{(3b)}}{A}\sim \left(2\,\Tilde{C}_3\,-\,1\right)\,n^2_s\qquad
\hbox{with}\qquad \Tilde{C}_3 =\frac{M_N}{g_S\,F_\pi}\,C +\,\frac{1}{2}\,C_\chi .
\end{equation}
Limiting ourselves to the pure L$\sigma$M case $C_\chi=0$, inducing $\tilde{C}_L=\tilde{C}_3$, the susceptibility $\chi_N$~\eqref{eq:xin} and the three-body repulsive contribution~\eqref{eq:THREEBOD} are directly related, as found in our previous works, e.g., Ref.~\cite{Ericson2007}. This constitutes a very important result linking chiral properties of the nucleon to the saturation mechanism. In the general case where $C_\chi\ne 0$, there is still a strong link between the susceptibility and the three-body repulsive contribution.

\subsection{Constraints from Lattice-QCD}

Those chiral properties of the nucleon, associated with explicit chiral symmetry breaking, namely the first and second derivatives of the nucleon mass with respect to the current quark mass, are thus very sensitive to the modeling of the nucleon. We have also   shown  that the scalar coupling constant, $g_S$, and the nucleon response parameter, $C_N$ (or $C$ or $\kappa_\mathrm{NS}$), depend on the quark substructure and the confinement mechanism as well as the effect of spontaneous chiral symmetry breaking. We will now show how they can be constrained  by lattice data. 

The nucleon mass, as well as other intrinsic properties of the nucleon (sigma term, chiral susceptibilities), are QCD quantities which are in principle obtainable from lattice simulations. The problem is that lattice calculations of this kind are still difficult for small quark masses, or equivalently small  pion mass ${\cal M}_\pi$. Here $\mathcal{M}_\pi$ represents the pion mass to leading order in the quark mass (i.e., ignoring the NLO chiral logarithm correction), $\mathcal{M}^2_\pi=2 m \,B=-2 m\,\langle\overline{q}\,q\rangle_{\chi L}/{F^2}$ (GOR relation). The quantities $F$ (the pion decay constant in the chiral limit) and $B$ are two low energy parameters appearing in chiral perturbation theory \cite{Leut2012}. In practice $\mathcal{M}_\pi$ deviates numerically very little from the bosonised NJL pion mass ${M}_\pi$. Typically at the time of the publication of the pioneering work from the Adelaïde group~\cite{LTY04} (that we will call hereafter AD1), these LQCD limitations were $m >50$~MeV and $\mathcal{M}^2_\pi>0.27$~GeV$^2$ (to be compared to the physical value, $0.02$~GeV$^2$). Hence  a technique was needed to extrapolate the lattice data to the physical region. The difficulty of the extrapolation is linked to the non analytical behaviour of the nucleon mass as a function of $m$ (or equivalently $\mathcal{M}^2_\pi$) which comes from the pion cloud contribution. The idea of the Adelaide group, \cite{LTY03,LTY04,TGLY04,AALTY10} (papers referred herafter as AD0, AD1, AD2 and AD3)  was to separate the pion cloud self-energy, $\Sigma_{\pi}(\mathcal{M}_{\pi}, \Lambda)$, from the rest of the nucleon mass and to calculate it with just  one adjustable cutoff parameter $\Lambda$ entering the form factor. Actually different cutoff forms for the pion loops (Gaussian, dipole, monopole, sharp) were used  with the adjustable parameter $\Lambda$. This formulation of Chiral Perturbation Theory (ChiPT) is thus called the Finite Range Regulator (FRR) method.  The remaining non pionic part is expanded in terms of powers of $\mathcal{M}^2_{\pi}$  as follows: 
\begin {equation}
M_N(\mathcal{M}^2_{\pi}) = 
a_{0}\,+\,a_{2}\,\mathcal{M}^2_{\pi}\, +\,a_{4}\,\mathcal{M}^4_{\pi}\,+ ...+\,\Sigma_{\pi}(\mathcal{M}_{\pi},\, \Lambda)
\label{eq:LATTICE}
\end{equation}
where $\Sigma_{\pi}(\mathcal{M}_{\pi}, \Lambda) = \Sigma^{(\pi)}(\mathcal{M}_{\pi},\, \Lambda)\,+\,\Sigma_\mathrm{tad}^{(\pi)} \,(\mathcal{M}_{\pi},\, \Lambda)$.

In AD1, which incorporates in the analysis  the effect of a tadpole contribution $\Sigma_\mathrm{tad}^{(\pi)} \,(\mathcal{M}_{\pi},\, \Lambda)$, the best-fit value for $a_2$ shows little sensitivity to the shape of the form factor, with a value $a_2\simeq 1.5$~GeV$^{-1}$, which corresponds to a non pionic piece of the light quark sigma commutator $\sigma_N^{(s)}=30$~MeV. In AD0 (which is actually the preprint version of AD1) and in the more recent paper, AD3, the contribution of the tadpole was not considered. Depending on  the precise method used in the lattice simulation, the preferred values for $a_2$ was  smaller, in the range $a_2\simeq 1.0$ to $1.2$~GeV$^{-1}$. Notice that taking  $a_2$  in the range $a_2\simeq 1.2$ to $1.5$~GeV$^{-1}$ corresponds to a non pionic piece of the light quark sigma commutator $\sigma_N^{(s)}=24$ to $30$~MeV.

In AD1, (which incorporates the effect of the pion tadpole) the best-fit value for $a_4$ shows again little sensitivity to the shape of the form factor, with a values $a_4\simeq -0.5$~GeV$^{-3}$. In AD0 and AD3,  depending on the precise method used in the lattice simulation, the preferred values for $a_4$ was even smaller, in the range $a_4\simeq - 0.2$ to $-0.25$~GeV$^{-3}$. \\

Ignoring the pion tadpole contribution to the nucleon mass, we assume that we can identify the pionic self-energy on the lattice with our model calculation described above. Consequently the first and second derivative of the non pionic piece of the lattice expansion, 
\begin{eqnarray}
Q_{S,L}^{(s)}&=&\frac{\partial M_N^\mathrm{(no\,pion)}}{\partial m}= \left(\frac{\mathcal{M}^2_\pi}{m}\right)  (a_2\,+\,a_4\,\mathcal{M}^2_\pi)\simeq
\left(\frac{\mathcal{M}^2_\pi}{m}\right)\,a_2\\
\chi_{N,L}^{(s)}&=&\frac{\partial^2 M_N^\mathrm{(no\,pion)}}{\partial m^2}= \left(\frac{\mathcal{M}^2_\pi}{m}\right)^2 \,(2\,a_4)
\end{eqnarray}
can be identified with the non pionic piece of the scalar charge, see Eq.~\eqref{eq:SCALQ}, and of the chiral susceptibility, see Eq.~\eqref{eq:CHINS}, derived above:
\begin{equation}
Q_S^{(s)} \equiv Q_{S,L}^{(s)}\,  \hbox{ and }
\chi_N^{(s)} \equiv \chi_{N,L}^{(s)} \, .
\end{equation}
One arrives at the important result:
\begin{equation}
a_2= \frac{F_\pi\, g_{S}}{M^2_{\sigma}}  
\end{equation}
\begin{equation}\label{eq:a2a4}
a_4 =-\frac{F_\pi\,g_{S}}{2 M^4_{\sigma }}\,\left(3\,-\,2\,\tilde{C}_L \right)\quad\hbox{with}\quad
\tilde{C}_L = \frac{M_N}{g_S\,F_\pi}\,C\,+\,\frac{3}{2}\,C_\chi .
\end{equation}
Our previous works \cite{Chanfray2007, Massot2008} coincide with these relations in the specific case of the L$\sigma$M effective potential ($C_\chi=0$). They provide two constraints on the parameters of the confining model. Also notice that the model results on the rhs of the above equations should be rigorously understood with the various parameters calculated in the chiral limit which are in practice very close to their values at the physical current quark mass.

 The very robust conclusion is that the lattice result is much smaller than the one obtained in a the simplistic linear sigma model ($C=C_\chi=0$), for which $a_4\simeq -3.5$~GeV$^{-3}$. Hence lattice data require a strong compensation from effects governing the three-body repulsive force needed for the saturation mechanism.

\begin{figure}
\centering
\includegraphics[width=0.8\textwidth,angle=0]{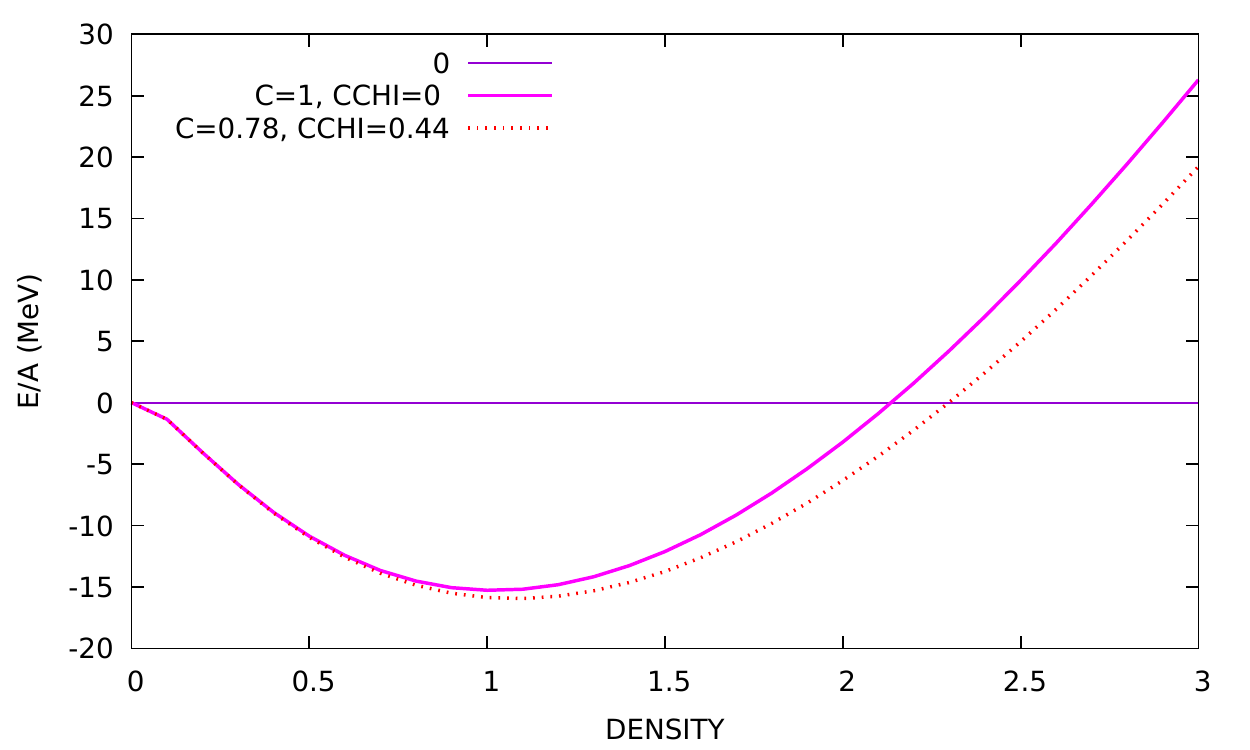}
\caption{Full line: original calculation of the EOS \cite{Chanfray2005} with $C=1,\, C_\chi=0$. Dotted line: New calculation with he same parameters but $C=0.78, \,C_\chi=0.44$. The density is scaled by the normal nuclear matter density.}
\label{fig:f3}
\end{figure}

\section{Discussion}\label{discussion}

The above results demonstrate that the lattice data $a_2$ and $a_4$, themselves related to the chiral responses of the nucleon, bring severe constraints on the nuclear matter equation of state. This suggests to enter these quantities as an input of a Bayesian analysis to generate the probability distribution function for the  nucleon response parameters $g_S$ and $C$. Such an analysis limited to the Hartree level has been performed in a recent work \cite{Rahul}, but using the simplistic L$\sigma$M, with an output for $C$ very close to $C\sim 1.5$, the obvious reason being the very small input value for $a_4$ (see Eq.(\ref{eq:a2a4})). In a work in preparation \cite{Cham}, we will perform again the same kind of analysis but with the incorporation of the Fock terms (and in particular the pion and rho Fock terms in presence of short range correlation) first with the L$\sigma$M and second with the enriched NJL chiral effective potential. As already mentioned, the problem of the analysis using the L$\sigma$M chiral effective potential is a large value of the $C$ response parameter in strong disagreement with all the nucleon models calculation which predict a value of $C$ smaller and most of the time significantly smaller than one (recall the MIT bag value $C\sim 0.5$). 

Just to have an insight on  the effect of an enriched chiral effective potential we return to our our original paper \cite{Chanfray2005}. In this paper where the L$\sigma$M was used we  obtained correct saturation properties with $C=1$  (see Fig. 1 of \cite{Chanfray2005}). We can retrospectively calculate the $a_2$ and $a_4$ parameters: we find $a_2=1.67$~GeV$^{-1}$ and $a_4=-1.48$~GeV$^{-3}$. If the obtained  $a_2$ is not very far from the lattice values, $a_4$ is in magnitude three times larger than the upper value compatible with lattice calculation. To see the effect of the NJL-like potential (via the parameter $C_\chi$), we simply incorporate the $(1-C_\chi)$ correction in the cubic term term of the L$\sigma$M chiral  effective potential, fixing $C_\chi=0.44$. Keeping all the other parameters at their original value, we take $C=0.78$, so as  to keep the same value of the repulsive three-body force, i.e., $ \Tilde{C}_3=C+C_\chi/2=1$ \eqref{eq:THREEBOD}. The saturation points is only slightly modified  (see Fig. \ref{fig:f3}) but now $\Tilde{C}_L=C+ 3C_\chi/2=1.44$ \eqref{eq:a2a4} and the $a_4$ parameter becomes very close to zero, $a_4=-0.1$~GeV$^{-3}$, in much better agreement with lattice data. 

\section{Conclusions}\label{conclusions}

The nuclear matter properties originate from the fundamental theory of the strong interaction and the aim of this manuscript is to investigate how this microscopic origin can be implemented in the modeling of nuclear matter. Of particular importance on the QCD side are the quark confinement mechanism and the chiral potential associated to the chirally broken QCD vacuum. 

In this article we use an enriched chiral effective potential, based on the NJL model, in place of the L$\sigma$M employed in  our previous phenomenological works. This significantly increases the agreement with LQCD data together with expected  model values of the nucleonic response parameter $C$. Note that this conclusion should be confirmed by a more thorough analysis (work in preparation \cite{Cham}). 

Hence the fundamental QCD theory and nuclear matter modeling are linked by, on the one hand the LQCD data $a_2$ and $a_4$ and on the other hand what we have called the "QCD connected parameters", namely the response parameters $G_s$ and $C_N$. Specifically we have shown that a particular combination of $C$ and $C_\chi$ ($\Tilde{C}_L$) is constrained by LQCD, which constitutes one of the main result of this paper. In addition a closely related combination ($\Tilde{C}_3=C+C_\chi/2)$) governs the repulsive  three-body force ensuring the mechanism mechanism.

Indeed these results provide a link between chiral properties of the nucleon and the saturation mechanism, already obtained in our previous works, but limited to the pure L$\sigma$M case.
Further investigations of these results shall be perform to understand more globally how they modify the properties of nuclear matter. Works in this direction is being performed.

\end{document}